\documentclass[review]{elsarticle}

\usepackage{amsfonts}
\usepackage{amsmath}
\usepackage{color}
\usepackage{epsfig}
\usepackage{graphics}
\usepackage{lineno,hyperref}
\modulolinenumbers[5]

\journal{Signal Processing}

\begin{document}

\begin{frontmatter}
\title{Complex Correntropy: Probabilistic Interpretation and Optimization  }

\author[UFRN]{Jo\~ao P. F. Guimar\~aes\corref{mycorrespondingauthor}}

\author[UFRN]{Aluisio I. R. Fontes}

\author[UFRN]{Joilson B. A. Rego}
\cortext[mycorrespondingauthor]{Corresponding author, Email address: joao.guimaraes@ifrn.edu.br}

\author[UFRN]{Allan de M. Martins}

\author[CNEL]{Jos\'e  C. Pr\'incipe }

\address[UFRN]{Federal University of Rio Grande do Norte, Brazil}
\address[CNEL]{Computational NeuroEngineering Lab - University of Florida, USA}

\begin{abstract}
	
Recent studies have demonstrated that correntropy is an efficient tool for analyzing higher-order statistical moments in nonGaussian noise environments. Although correntropy has been used with complex data, no theoretical study was pursued to elucidate its properties, nor how to best use it for optimization . This paper presents a probabilistic interpretation for correntropy using complex-valued data called complex correntropy. A recursive solution for the maximum complex correntropy criterion (MCCC) is introduced based on a fixed-point solution. This technique is applied to a simple system identification case study, and the results demonstrate prominent advantages when compared to the complex recursive least squares (RLS) algorithm. By using such probabilistic interpretation, correntropy can be applied to solve several problems involving complex data in a more straightforward way.

\end{abstract}


\end{frontmatter}

\linenumbers

\section{Introduction}




Defining the relationship between the input and output signals in a given system is a common problem widely found in distinct engineering areas \cite{exemplo1,exemplo2,exemplo3,voz}. The classic regression solution is extensively adopted using the mean square error (MSE) as a cost function in order to minimize the error between the input signal and the desired output. However, many authors have developed methods based on correntropy as a cost function in the last few years since such approach improves the fitting performance in nonGaussian noise environments \cite{propriedadesCorrentropia,corretropia2,Fontes2015}.

Correntropy is a similarity measure between two variables, which contains a weighted combination of all the even statistical moments, being a generalization of the correlation concept \cite{Santamaria2006a}. Several works have proposed the use of correntropy in adaptive system training achieving excellent performance in  practical applications where the errors are typically nonGaussian \cite{xu2008pitch,linhares2015fuzzy,liu2007correntropy,liu2013correntropy,rego2linear,he2011maximum}.

On the other hand, many applications involves signal sources that are defined in the complex domain. Statistical signal processing in the complex domain has traditionally been viewed as a straightforward extension of the corresponding algorithms in the real domain~\cite{alunomandic} such as the complex RLS \cite{livroRLSComplexo}. One may notice that the complex algorithms sometimes look similar to their variants for real-valued data but this is not always the case when nonlinearities are involved.

Few studies have explored the use of correntropy as a cost function in problems involving complex-valued data. The work developed in \cite{koreano} presents a complex-valued blind equalization algorithm for quadrature amplitude modulation (QAM) and complex channel environments based on the correntropy criterion. The study is motivated by the improved performance achieved by information theoretic learning (ITL) methods when compared to MSE-based approaches. A robust adaptive carrier frequency offset (CFO) algorithm was introduced in \cite{iraniano} for orthogonal frequency division multiplexing (OFDM) purposes, which also deals with QAM and phase-shift keying (PSK) complex symbols. However, correntropy as applied to complex-valued data has not yet been properly formalized.

This paper presents a probabilistic interpretation for correntropy with complex-valued data, which is defined as complex correntropy. This probabilistic interpretation provides further insights and is based on the probability function in multidimensional spaces using the Parzen estimator. In addition, for optimization of system parameters, the maximum complex correntropy criterion (MCCC), is used as a cost function in a system identification problem with complex-valued data. As the MCCC is a real-valued function but depends on a complex-valued parameter is not analytic. Therefore, standard differentiability does not apply because the Cauchy-Riemann conditions are violated. Thus, it is necessary to employ the Wirtinger derivatives which are based on the duality between the spaces $\mathbb{C}$ and $\mathbb{R}^{2}$\cite{mandic2009complex} to obtain a recursive solution based on a fixed-point algorithm. The results demonstrate the advantages of the proposed cost function in nonGaussian environments when compared to the Complex RLS algorithm for noise cancellation purposes.

The remaining sections of this paper are organized as follows: Section II reviews correntropy and extends its concept to complex-valued data. Section III presents a closed form recursive solution to MCCC. Simulation results are presented in section IV, while a proper comparative analysis with the RLS algorithm performance is presented. Finally, conclusions are given in Section V.

\section{ Probabilistic Interpretation of Correntropy }
\label{PI}

This section reviews the probabilistic interpretation of correntropy applied to real-valued data, so that it can be further extended to complex-valued data.

\subsection{Correntropy Applied to Real-Valued Data}
\label{CRN}

Correntropy is directly related to the estimation of how similar two random variables are when a Parzen estimator is used for the joint probability \cite{livroitl}. Firstly, let us consider two arbitrary scalar random variables $\textit{X}$ and $\textit{Y}$ with a smooth joint probability density function $f(x,y)$. The probability density of the event $X=Y$ can be written as

\begin{equation}\label{c1}
P(X = Y) = \int_{-\infty}^{\infty}  \int_{-\infty}^{\infty} f_{XY}(x,y) \delta(x-y) \mathrm{d}x\mathrm{d}y
\end{equation}

In most cases, the real distribution is unknown and only a finite number of data samples ${(x_{n},y_{n}), n= 1,2,...N}$ is available. However, it is possible to use the L-dimensional Parzen estimation with a Gaussian kernel to obtain the estimate of the joint PDF $f_{XY}(x,y)$ \cite{silverman1986density} as:

\begin{equation}\label{parzenL}
\hat{f}_{X^{1},X^{2},...X^{L}}(x^{1},x^{2},...,x^{L}) = \frac{1}{N}\sum\limits_{n=1}^N \prod \limits_{l=1}^L  G_{\sigma}(x^{l}-x^{l}_{n})
\end{equation}
where $G_{\sigma}(x)$ is defined as

\begin{equation}\label{kernel}
G_{\sigma}(x) = \frac{1}{\sqrt{2\pi}\sigma}exp \left ( -\frac{x^2}{2\sigma^2} \right )
\end{equation}

Notation $x^{l}_{n}$ represents the $n$-th data sample for the $l$-th component of the L-dimensional random vector while $\sigma$ is the kernel bandwidth, also known as the kernel size. In order to define correntropy in the real domain, the work presented in \cite{integralxy} considers L=2 (making $X^1=X$ and $X^2=Y$) in equation (\ref{parzenL}):

\begin{equation}\label{Parzenclassica}
\hat{f}_{XY}(x,y)= \frac{1}{N} \sum\limits_{n=1}^N  G_{\sigma}( x - x_{n}) G_{\sigma}(y - y_{n})
\end{equation}

Substituting (\ref{Parzenclassica}) in (\ref{c1}) gives:

\begin{equation}\label{c4}
\hat{P}(X=Y) = \int_{-\infty}^{\infty}\int_{-\infty}^{\infty}  \frac{1}{N} \sum\limits_{n=1}^N  G_{\sigma}(x - x_{n})G_{\sigma}(y - y_{n}) \delta(x-y) \mathrm{d}x\mathrm{d}y
\end{equation}

Since the only nonzero values occur along the bisector of the joint space (because of the delta function), $x=y$, equation (\ref{c4}) can be rewritten as:

\begin{equation}\nonumber
\hat{P}(X=Y) = \int_{-\infty}^{\infty}  \frac{1}{N} \sum\limits_{n=1}^N G_{\sigma}(x - x_{n})G_{\sigma}(y - y_{n}) \mathrm{d}u \Big|_{x=y=u} 
\end{equation}

\begin{equation}\label{c2}
\hat{P}(X=Y) = \int_{-\infty}^{\infty}  \frac{1}{N} \sum\limits_{n=1}^N  G_{\sigma}( u - x_{n})  G_{\sigma}(u - y_{n}) \mathrm{d}u 
\end{equation}
where $u$ represents the value assumed by $x$ and $y$ over the line $x=y$.

Because the integral of the product of Gaussians is a Gaussian with a kernel size equal to the square root the original, Equation (\ref{c2}) can be written as

\begin{equation}\label{classica}
\hat{P}(X=Y) = \frac{1}{N}\sum\limits_{n=1}^N G_{\sqrt{2}\sigma}(x_{n}-y_{n})
\end{equation}

Recall that correntropy can be estimated as $V(X,Y) = E_{XY}[G_{\sqrt{2}\sigma}(X-Y)]$ using the Gaussian kernel. Hence we can write 

\begin{equation}\label{eq:estimative}
	V(X,Y) = \int_{-\infty}^{\infty}\int_{-\infty}^{\infty} f_{XY}(x,y)G(x-y)\mathrm{d}x\mathrm{d}y
\end{equation}

Therefore, for smooth pdfs, correntropy can indeed be interpreted in the limit of small kernel size as the density of the event $X=Y$ for two random variables. However, in a non parametric estimation from samples using Parzen windows, correntropy can be estimated by \ref{classica} for $X$ and $Y$ real random variables for any finite kernel size \cite{livroitl}. This explains in simple terms the difference between the correntropy criterion versus the mean square error, which only quantify second order moments.

\subsection{Correntropy Applied to Complex-Valued Data}
\label{CCN}

This paper presents a probabilistic interpretation based on Parzen estimator defined according to equation (2) to measure the similarity between two complex variables. We will basically use the methodology developed in the previous section to extend correntropy to more than two variables, preserving the probability interpretation. Assuming two random complex variables $C_{1}=X+j\,Z$ and  $C_{2} = Y+j\,S$, where $C_{1},C_{2} \in \mathbb{C}$, and $X,Y,Z,S$  are real-valued random variables, we can estimate the  probability density of the event $C1=C2$ as

\begin{equation}\label{integral4}
\hat{P}(C_{1} = C_{2}) = \int_{-\infty}^{\infty} \int_{-\infty}^{\infty} \int_{-\infty}^{\infty} \int_{-\infty}^{\infty} \hat{f}_{XYZS}(x,y,z,s) \delta(x-y) \delta(z-s) \mathrm{d}x\mathrm{d}y \mathrm{d}z\mathrm{d}s
\end{equation}
If $x=y$ and $z=s$, equation (\ref{integral4}) can be rewritten as:

\begin{equation}\nonumber
\hat{P}(C_{1} = C_{2}) =\int_{-\infty}^{\infty} \int_{-\infty}^{\infty}  \! \hat{f}_{XYZS}(x,y,z,s) \, \mathrm{d}u_{1}\mathrm{d}u_{2} \Big|_{x=y=u_{1}, z=s=u_{2}} 
\end{equation}

\begin{equation}\label{integraldupla}
\hat{P}(C_{1} = C_{2}) =\int_{-\infty}^{\infty} \int_{-\infty}^{\infty}  \! \hat{f}_{XYZS}(u_{1},u_{1},u_{2},u_{2}) \, \mathrm{d}u_{1}\mathrm{d}u_{2} 
\end{equation}

It is then possible to replace $\hat{f}_{XYZS}$ for the Parzen estimator defined in equation (\ref{parzenL}) using  $L = 4$:

\begin{equation}\nonumber
=\int_{-\infty}^{\infty} \int_{-\infty}^{\infty} \frac{1}{N}\sum\limits_{n=1}^N  G_{\sigma}(x-x_{n}) G_{\sigma}(y-y_{n})\,G_{\sigma}(z-z_{n}) G_{\sigma}(s-s_{n})  \mathrm{d}u_{1} \mathrm{d}u_{2} \Big|_{x=y=u_{1},z=s=u_{2}}
\end{equation}


\begin{equation}\label{dupla}
=\frac{1}{N}\sum\limits_{n=1}^N  \int_{-\infty}^{\infty} \int_{-\infty}^{\infty} G_{\sigma}(u_{1}-x_{n}) \, G_{\sigma}(u_{1}-y_{n}) \,G_{\sigma}(u_{2}-z_{n}) \, G_{\sigma}(u_{2}-s_{n})  \mathrm{d}u_{1} \mathrm{d}u_{2} 
\end{equation}

Solving the double integral in (\ref{dupla}) gives:

\begin{equation}\label{final}
\hat{P}(C_{1} = C_{2}) = \frac{1}{N}\sum\limits_{n=1}^N  G_{\sigma \sqrt{2}}(x_{n} - y_{n}) \,G_{\sigma \sqrt{2} }(z_{n} - s_{n}) 
\end{equation}

Using the previous argument, this is also the estimate of the correntropy for two complex random variables $C_{1}$ and $C_{2}$. Combining the inner exponential terms we can write the product of $(C_{1} - C_{2})$ by its conjugate $(C_{1} - C_{2})^{*}$ pondered by the kernel size $2\sigma^{2}$ as

\begin{equation}\nonumber
(C1-C2)\,(C1-C2)^{*} = (X - Y)^{2} + (Z - S)^{2}  
\end{equation}

Hence, correntropy for two complex random variables or simply complex correntropy will then be defined as 

\begin{equation}\label{finalprincipe}
V^{C}(C_{1},C_{2}) =  E_{C_{1}C_{2}}[ G^{C}_{\sigma\sqrt{2}} (C_{1} - C_{2}) ] 
\end{equation}
where 
\begin{equation}
G^{C}_{\sigma} (C_1 - C_2 )= \frac{1}{2\pi\sigma^2}exp \left ( -\frac{(C_{1} - C_{2}) (C_{1} - C_{2})^{*}}{2\sigma^2} \right )
\end{equation}
and $*$ means the complex conjugate. 

There are no assumptions or restrictions for its application to generic experimental methods e.g. constant modulus or argument, since  it represents a complete measure of similarity between two complex random variables. The non parametric estimator of complex correntropy with Parzen windows can be written as

\begin{equation}
V^{C}(C_{1},C_{2}) = \frac{1}{2\pi\sigma^{2}} \frac{1}{N} \sum\limits_{n=1}^N exp \left ( -\frac{(x_{n} - y_{n})^{2} + (z_{n} - s_{n})^{2} }{2\sigma^2} \right ) 
\end{equation}






It is important to understand the effect of the estimator when computing correntropy as a probability density estimation. There is an inherent compromise in selecting the bandwidth of the Gaussian kernel, because on the one hand the kernel should emphasize samples in the bisector of the joint space (small kernel size), but on the other, consider the effect of as many samples as possible (large kernel). 

Equation (\ref{final}) can also be further analyzed according to its respective Taylor series expansion. In addition, it is possible to write the average sum as the expected value in the Parzen estimator, which leads to:

\begin{equation}\nonumber
V^{C}(C_{1},C_{2}) = \frac{1}{2 \pi \sigma^2} \sum\limits_{m=0}^\infty  \dfrac{(-1)^m}{2^m \sigma^{2m} n! }E_{XY}[(X-Y)^{2m} + (Z-S)^{2m}]
\end{equation}

\begin{equation}\label{taylor}
V^{C}(C_{1},C_{2}) = \frac{1}{2 \pi \sigma^2} + \frac{k_{1}}{\sigma^4}E_{XY}[(C_{1} - C_{2})(C_{1} - C_{2})^{*}] + h_{\sigma^6}(C_{1} - C_{2}) 
\end{equation}
where $h_{\sigma^6}(C_1- C_2)$  is a term that contains all higher-order moments, whose components in the denominator depend on $\sigma$ considering that the first term includes $\sigma^{6}$. 

According to equation (\ref{taylor}), the higher-order terms represented by $h_{\sigma^6}$ tend to zero faster than the second term as $\sigma$ increases. It is worth mentioning that the second term corresponds exactly to the covariance involving two complex variables $C_{1}$ and $C_{2}$. Hence, as the kernel size increases, the complex correntropy tends to the covariance analogously to the real value case.

\section{Maximum Complex Correntropy Criterion (MCCC)}
\label{FPS}


A typical system identification task is represented in Fig. 1. Since correntropy has been previously defined in the complex domain, it is necessary to establish the MCCC. 

Let us consider a linear model and define the error $e$ as being the difference between the desired signal $d$ and the filter output 
$y$ where $\textbf{x}$, $\textbf{w}$, $y$, $d$, $e$  $\in \mathbb{C}$. Then

\begin{equation}\label{relacao1}
y = \textbf{w}^H \textbf{x} \quad \text{and} \quad e = d-y
\end{equation}

Let the new criteria MCCC be defined as the maximum complex correntropy between two random complex variables $D$ and $Y=\textbf{w}^H\,\textbf{X}$. 


\begin{equation}\label{CCJ3}
J_{MCCC} = V^{C}(D, Y) = E_{DY}[G^{C}_{\sigma\,\sqrt{2}}(D-\textbf{w}^{H}\textbf{X})]
\end{equation}

The fixed-point solution for the optimal weights can be obtained by setting the cost function derivative to zero in respect to $\textbf{w}^*$ in equation (\ref{CCJ3}). But, as mention previously, equation (\ref{CCJ3}) is not an analytical function in the complex domain. Thus, it is necessary to use the Wirtinger Calculus \cite{mandic2009complex} to compute its derivative, which yields,



\begin{equation}
E_{DY}[G^{C}_{\sigma\,\sqrt{2}}(e) \textbf{X}(D^* - \textbf{w}^{T}\textbf{X}^{*})] = \textbf{0}
\end{equation}

\begin{equation}\label{deduzindo_passo3}
E_{DY}[G^{C}_{\sigma\,\sqrt{2}}(e)\textbf{X}\,D^{*}] = E_{DX}[G^{C}_{\sigma\,\sqrt{2}}(e)\textbf{XX}^{H}]\,\textbf{w}
\end{equation}

Notice that although \ref{deduzindo_passo3} has the same functional form as the Wiener solution, it is not an analytic solution because of the inclusion of the error in each side of the equation, which is a function of the filter parameter W. However it can be estimated as a fixed point equation as 

\begin{equation}\label{pontofixo}
\textbf{w} = \left [ \sum_{n=1}^{N} G^{C}_{\sigma\,\sqrt{2}}(e_{n})\textbf{x}_{n}\textbf{x}_{n}^{H} \right ]^{-1} \left [ \sum_{n=1}^{N} G^{C}_{\sigma\,\sqrt{2}}(e_{n})\textit{d}_{n}^{*}\,\textbf{x}_{n} \right ]
\end{equation}
which represent an iterative solution to obtain the optimal value of $\textbf{w}$. Even though convergence is achieved after a few iterations, each one of them requires the computation of the whole sum, which is inadequate for real-time learning. A fixed-point stochastic recursive solution can then be derived as inspired by \cite{singh2010closed} and based on equation (\ref{pontofixo}). Firstly, let us  define a weighted auto correlation matrix of the complex input signal, and a weighted cross correlation vector between the desired conjugate and the input vector as:

\begin{equation}\label{eqs}
R_{n} = \sum_{n=1}^{N} G^{C}_{\sigma\,\sqrt{2}}(e)\textbf{x}_{n}\textbf{x}_{n}^{H}
\quad \text{and} \quad
P_{n} = \sum_{n=1}^{N} G_{\sigma\,\sqrt{2}}(e)\textit{d}_{n}^{*}\,\textbf{x}_{n}
\end{equation}

As \cite{singh2010closed} showed for the real-value case, the equations (\ref{eqs}) resemble to the Wiener solution but instead of using the simple average, the exponential Gaussian function of the error is used to weight the average. The auto correlation matrix and the cross correlation vector can be updated recursively such as in the classical least square or RLS algorithm \cite{haykin2002adaptive}, thus we obtain

\begin{equation}\label{eqr2}
R_{n} = R_{n-1} + G^{C}_{\sigma\,\sqrt{2}}(e)\textbf{x}_{n}\textbf{x}_{n}^{H}
 \quad \text{and} \quad P_{n} = P_{n-1} + G_{\sigma\,\sqrt{2}}(e)d_{n}^{*}\,\textbf{x}_{n}
\end{equation}

In order to implement the recursive expressions represented by \ref{eqr2}, it is necessary to pick an initial $w$ parameters as well as an initial values for $R_{0}$ and $P_{0}$.

\section{Results}

In order to evaluate the MCCC performance, the complex RLS algorithm presented in \cite{livroRLSComplexo} has been adopted for comparison purposes. Besides, the weight signal-to-noise ratio (WSRN) is also considered in the analysis of results as in \cite{singh2010closed}, since it quantifies convergence and misadjustment rates properly in decibels as:

\begin{equation}
WSNR_{db} = 10 \log_{10} \left (  \frac{ \bar{w}^{H}\,\bar{w}  }{ (\bar{w} - w_{n})^{H}(\bar{w} - w_{n}) }  \right )
\end{equation}
where $\bar{w} = [(+1-2j),(-3+4j)]^{T}$ is the proper weight chosen for the simulation tests and $w_{n}$ is the weight computed by the aforementioned methods in the $n$-th iteration.


The desired signal is formed by the product of the input signal $X = [X_{1} X_{2}]$ and $\bar{w}$. $X_{1}$ and $X_{2}$ are both random complex variables with PDF (probability density function) $\mathcal{N}(0.0,1.0)$, where $\mathcal{N}(\mu,\sigma)$ is a normal Gaussian distribution with mean $\mu$ and variance $\sigma^2$. Then, this signal was contaminated with nonGaussian noise whose PDF is $0.95\mathcal{N}(0.0,0.05) + 0.05\mathcal{N}(0.0,5.0)$. The authors in \cite{singh2010closed} also employ the aforementioned PDF to represent the noise and evaluate robustness of the fixed-point MCC algorithm compared with its RLS counterpart, although data only comprises the real domain. The noise signal $\eta_{n}$ is then generated in this work, where $\eta \in \mathbb{C}$ and  $\eta_{n} = \eta^{re}_{n} + j\,\eta^{im}_{n}$, while $\eta^{re}_{n}$ and $\eta^{im}_{n} \in \mathbb{R}$ and follow the described PDF.

After 300 iterations, the results shown in Fig. 2 could be obtained. The curves represent the average when using 50 Monte Carlo trials, as weights always start from zero. It can be stated that the proposed approach is able to ignore outliers specially with $\sigma = 1$. The kernel size in equation (\ref{taylor}) behaves as a parameter that weights both second-order $(m=1)$ and higher-order moments. As $\sigma$ becomes higher than unity, the high-order moments decrease faster as the achieved results are closer to the ones provided by the conventional complex RLS solution.

\section{Conclusions}

This paper has presented the extension of the correntropy concept to complex-valued data in an approach defined as complex correntropy. A significant contribution of this work lies in obtaining the expression for the complex correntropy from its respective probabilistic interpretation. Besides, a recursive algorithm based on fixed-point solution has been introduced, which can be used to derive the MCCC. Simulation tests have also demonstrated that the proposed method presents high convergence rates, but with higher efficiency when dealing with outlier environments if compared to the complex RLS approach. It is then reasonable to state that correntropy can now be applied to the solution of distinct problems involving complex data in a more straightforward way. 

\section*{References}

\bibliography{mybibfile}

\newpage
\begin{figure}[h!]
	\centering
	\includegraphics[width=3.5in]{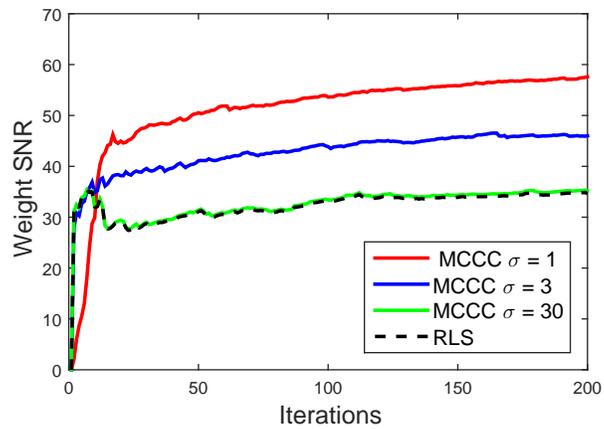}
	\caption{Weight SNR plots for MCCC fixed point  } 
	\label{wsnr}
\end{figure}

\begin{figure}[h!]
	\centering
	\input{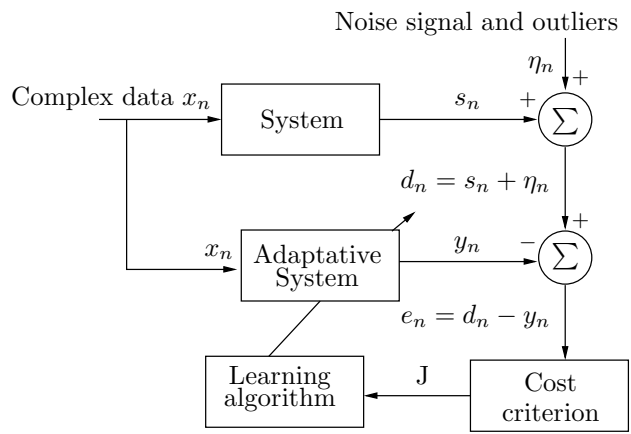}
	\caption{Typical system identification}
	\label{figuraSistema}
\end{figure}

\end{document}